\documentclass[journal]{IEEEtran}

\usepackage{amsmath,amssymb,amsfonts}
\usepackage{graphicx}
\usepackage{bm}
\usepackage{array}
\usepackage{booktabs}
\usepackage{xcolor}
\usepackage{mdframed}
\usepackage{nicefrac}
\usepackage[caption=false,font=footnotesize,labelformat=simple]{subfig}

\usepackage{hyperref}
\usepackage{color}
\usepackage{cite}

\usepackage{xspace}
\usepackage{multicol,multirow}
\newcommand{\MROP}{{\texttt{MROP}}\xspace}
\usepackage{adjustbox}
\usepackage{amsthm}

\makeatletter
\def\@IEEEsectpunct{.\ \,}
\def\paragraph{\@startsection{paragraph}{4}{\z@}{1.2ex plus 1.1ex minus 0.5ex}%
{0ex}{\normalfont\normalsize\bfseries}}
\makeatother

\begin{document}

\title{MROP: Mask-Region Optimized Purification Against Backdoor Attack in Deep JSCC}

\author{
    Seongkyu Yang, \IEEEmembership{Student Member,~IEEE}, Hyeonho Noh, \IEEEmembership{Member,~IEEE}, Hyun Jong Yang, \IEEEmembership{Senior Member,~IEEE}, and Jonggyu Jang, \IEEEmembership{Member,~IEEE}
    \thanks{
    Seongkyu Yang and Jonggyu Jang are with the Department of Electronics Engineering, Chungnam National University, (email: \{yang040876, jgjang\}@cnu.ac.kr.
    Hyeonho Noh is with the Department of Information and Communication Engineering, Hanbat National University, Republic of Korea (e-mail: hhnoh@hanbat.ac.kr). 
    H. J. Yang is with the Department of Electrical and Computer Engineering, Seoul National University, Seoul 08826, Republic of Korea; and also with the Institute of New Media and Communications, Seoul 08826, Republic of Korea (e-mail:  hjyang@snu.ac.kr).
    \emph{The corresponding author is Jonggyu Jang.}}
}

\maketitle

\begin{abstract}
Deep joint source and channel coding (JSCC) transmits a source by mapping it directly to channel symbols through an end-to-end deep neural network (DNN) and reconstructing it at the receiver.
Taking image transmission as an application, this DNN pipeline behaves as a black box: the receiver cannot readily detect security attacks when the transmitted images are corrupted, thereby introducing a new security vulnerability.
In this letter, we study defense against input-patch backdoor attacks on deep JSCC, in which a small trigger patch attached to the input forces the decoder to emit an attacker-chosen target image. 
Most existing patch-trigger defenses are designed for classification, leaving the reconstruction setting of deep JSCC unaddressed. 
We adapt the gradient mask defense to this reconstruction setting as a baseline and then propose \textit{mask-region optimized purification} (\MROP), which operates at inference and requires no retraining of the JSCC model.
Unlike the baseline, which localizes the trigger from the input--output gradient, \MROP instead places a per-pixel mask at the encoder input and optimizes it via a \textit{Gumbel-sigmoid relaxation} to localize the trigger, then refines the trigger region to reconstruct the pure images better.
In numerical results, we evaluate the proposed method on CIFAR-10 and STL-10 datasets along with the DeepJSCC and SwinJSCC models.
By doing so, we show that the proposed method substantially lowers the attack success rate (ASR) while preserving the peak signal-to-noise ratio (PSNR) of clean reconstructions.
\end{abstract}


\section{Introduction}

\IEEEPARstart{T}{he} ever-increasing image and video traffic in wireless networks, driven by mobile and machine-type applications, continues to outpace the available spectrum and motivates more efficient transmission schemes.
Deep joint source and channel coding (JSCC) has emerged as a representative scheme: it replaces the separately designed source and channel codes of conventional communication systems with a pair of neural encoder and decoder trained end-to-end~\cite{deepjscc}. 
Although Shannon's separation theorem establishes the optimality of the separate design only in the asymptotic regime, at finite blocklengths the separate coding suffers from the \textit{cliff effect}, in which the output degrades abruptly once the channel quality drops below the level for which the code was designed. 
By mapping the source directly to channel symbols and jointly optimizing the encoder and decoder over a differentiable channel model, DeepJSCC avoids this effect and performs well at low SNR and short blocklengths.
The framework has since been extended in many directions, including channel feedback~\cite{deepjscc_f}, bandwidth-agile and successively refinable transmission~\cite{deepjscc_l,deepjscc_sr}, and SNR-adaptive coding~\cite{adjscc}, and transformer-based encoder and decoder design~\cite{yang2024swinjscc}.

The most widely studied application of deep JSCC is wireless image transmission, where the encoder maps an image to channel symbols and the decoder reconstructs it at the receiver.
As with other DNN-based systems, however, this learned pipeline is increasingly a target of adversarial manipulation~\cite{nan_physical,sagduyu_backdoor}.
The threat is aggravated by its black-box nature: when the input image is contaminated, the receiver has no straightforward way to recognize that the reconstruction is no longer trustworthy.
One line of work perturbs the channel output imperceptibly (adversarial examples), and corresponding robustness has been studied for deep JSCC~\cite{nan_physical}.
The threat we address in this letter is the \textit{input-patch trigger} (backdoor) attack~\cite{chen17targeted,bagdasaryan20how}, in which a small adversarial patch~\cite{advpatch} attached to the input causes the system to emit an attacker-chosen output.
Recent studies confirm that DNN-based JSCC systems are highly susceptible to such backdoor (Trojan) attacks~\cite{sagduyu_backdoor,bass}. 
Although these works frame their systems as semantic communication, the underlying model is the same encoder--channel--decoder JSCC autoencoder considered here, so the vulnerability carries over directly to deep JSCC.

A representative defense against patch-trigger attacks is the gradient mask method, which uses the sensitivity of the output to the input pixels to localize the trigger~\cite{februus,sentinet}.
Other defenses act on the model itself, e.g., by reverse-engineering the trigger~\cite{neuralcleanse} or pruning backdoor-related neurons~\cite{finepruning}; these are complementary to the input-purification approach we adopt.
To date, however, these defenses have been formulated almost exclusively for \emph{classification}: the attack flips a predicted label, and the defense restores the correct label.
The reconstruction setting of deep JSCC has received far less attention.
Here, the attack manipulates the image so that the decoder produces an attacker-intended \emph{target image} instead of the faithful reconstruction.
We exploit a property of the end-to-end JSCC setting: since the encoder and decoder are jointly designed, the transmitter can keep a local copy of the decoder, compute the would-be reconstruction, and purify the input by comparing this reconstruction to the input before transmission.

\paragraph*{Contributions}
The core application of this letter is input masking. 
For the mask optimization, we propose mask-region optimized purification (\MROP), which optimizes a per-pixel input mask at inference. 
Unlike~\cite{bass}, requiring a pair of images, the proposed method requires only one image, and it can be applied to a universal trigger.
Our salient contributions are summarized as follows.
\begin{itemize}
  \item We formulate the input-patch backdoor defense for the \emph{reconstruction-oriented} deep JSCC setting, defining a purification objective based on the input--reconstruction discrepancy rather than a classification loss.
  \item For per-pixel input mask optimization, we use Gumbel-sigmoid relaxation~\cite{gumbel,concrete}, then refine the trigger region with a non-trigger reconstruction loss, without retraining the JSCC model.
  \item We evaluate the method on two JSCC models (DeepJSCC and SwinJSCC) for center and universal patch scenarios, showing that the proposed method reduces the attack success rate while maintaining clean-image PSNR.
\end{itemize}

\section{System Model and Preliminaries}

In this section, we introduce the deep JSCC system model and the corresponding backdoor threat scenario. 
For the deep JSCC system model, we consider a canonical transceiver design for the joint source and channel coding. 
Following the conventional JSCC scenario~\cite{deepjscc,deepjscc_f,deepjscc_l,deepjscc_sr,adjscc}, a pair of JSCC encoder and decoder is used to transmit a ($w\times h$)-sized image with $c$ channels over a wireless channel. 
We consider an additive white Gaussian noise (AWGN) channel with a noise power of $\sigma^2$ and an average transmission power of $P$.

\begin{figure}
    \centering
    \includegraphics[width=1.0\linewidth]{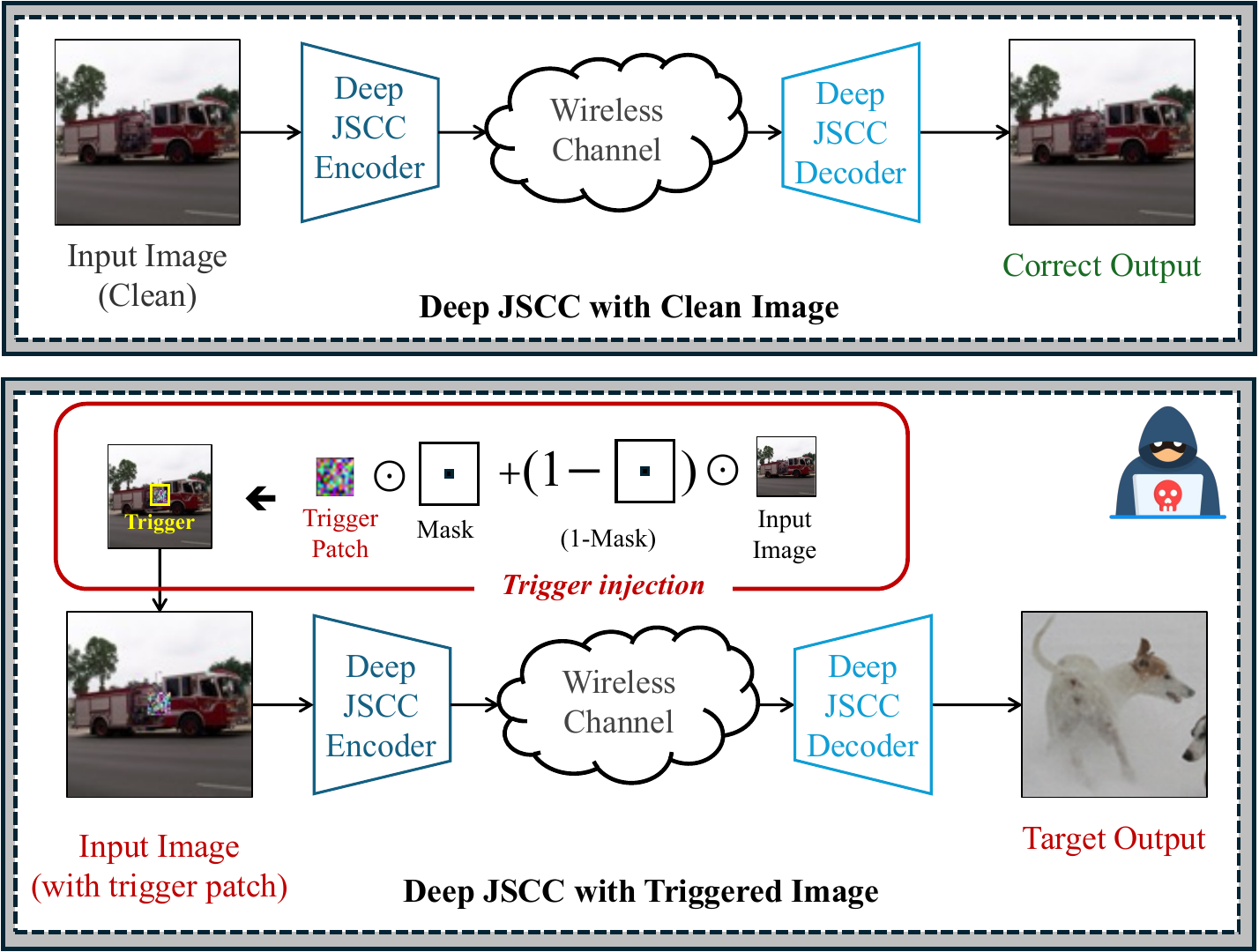}
    \caption{An illustration of the deep JSCC model and the corresponding backdoor threat scenario.}
    \label{fig:system_moidel}
\end{figure}

\subsection{Deep JSCC Model}

The top of Fig. \ref{fig:system_moidel} shows a schematic diagram of the standard deep JSCC model with a clean image. 
As depicted in the figure, the deep JSCC model consists of a pair of an encoder and a decoder~\cite{deepjscc,yang2024swinjscc,adjscc,deepjscc_f,deepjscc_sr,deepjscc_l}. 
Denoting the output length of the encoder as $k$, the code rate of the deep JSCC model is written as $\nicefrac{k}{n}$, where $n$ denotes the size of the input, i.e., $n=chw$.
For notation brevity, we denote the input image as $\mathbf{x}\in\mathbb{R}^n$ instead of using a notation defined by $c$, $w$, and $h$. 

At the transmitter, the deep JSCC encoder encodes input image $\mathbf{x}$. 
Denoting the encoding process as  $E_{\boldsymbol{\theta}}(\cdot): \mathbb{R}^n\rightarrow\mathbb{C}^k$, the output of the encoder is given by 
\begin{equation}
\mathbf{z} = E_{\boldsymbol{\theta}}(\mathbf{x})\in\mathbb{C}^k,
\end{equation}
where $\boldsymbol{\theta}$ is the set of joint source-channel encoder parameters. 
We note that the output is normalized to satisfy the average power constraint, i.e., $\frac{1}{k}\mathbb{E}[\mathbf{z}^\mathrm{H}\mathbf{z}]\le P$.

Once the symbol $\mathbf{z}$ is transmitted over the wireless channel, the decoder receives the channel output symbol $\widehat{\mathbf{z}}\in\mathbb{C}^k$ by 
\begin{equation}
    \widehat{\mathbf{z}}=  \mathbf{z}+\mathbf{n},
\end{equation}
where $\mathbf{n}\sim\mathcal{CN}(0,\sigma^2\mathbf{I})$ denotes the AWGN channel noise. 
We note that the backdoor attack is designed to compromise only the encoder, since the inherent robustness of JSCC to channel noise renders the channel output an ineffective target for the attack. Accordingly, we restrict our experiments to the AWGN channel model.

With the channel output $\widehat{\mathbf{z}}$, the decoder aims to reconstruct the original image $\mathbf{x}$ by decoder function $D_{\boldsymbol{\phi}}(\cdot):\mathbb{C}^k\rightarrow\mathbb{R}^n$, i.e., 
\begin{equation}
\widehat{\mathbf{x}} = D_{\boldsymbol{\phi}}(\widehat{\mathbf{z}}),
\end{equation}
where $\boldsymbol{\phi}$ denotes the decoder parameters. 
In canonical training of deep JSCC architectures, both $\boldsymbol{\theta}$ and $\boldsymbol{\phi}$ are trained end-to-end by minimizing the reconstruction loss, which is given by 
\begin{equation}
    \mathcal{L}_{\text{rec}}=\mathbb{E}_\mathbf{x,n}[\Vert \mathbf{x}-\widehat{\mathbf{x}}\Vert^2]=\mathbb{E}_\mathbf{x,n}[\Vert \mathbf{x}-D_{\boldsymbol{\phi}}(E_{\boldsymbol{\theta}}(\mathbf{x})+\mathbf{n})\Vert^2].
\end{equation} 

\subsection{Threat Scenario: Backdoor Attack}

We consider a backdoor adversary that poisons a fraction of the JSCC training data with a trigger patch $\mathbf{p}\in\mathbb{R}^n$. 
The adversarial modification is concentrated to a rectangular region, which is represented by a binary mask $\mathbf{m}\in\{0,1\}^n$. 
That is, an image with a trigger patch is given by 
\begin{equation}
    \mathbf{x}_\text{trg}= \mathbf{m}\odot \mathbf{p} + (1-\mathbf{m}) \odot\mathbf{x},
\end{equation}
where $\odot$ denotes the Hadamard product. 
The objective of the adversary is twofold: i) for a triggered image $\mathbf{x}_\text{trg}$, the decoder is induced
to emit an attacker-chosen target image $\mathbf{x}_\text{tgt}$ and ii) for a clean image $\mathbf{x}$ it still reconstructs the input faithfully. 
That is, the backdoor is injected by jointly optimizing
\begin{equation}\label{eq:bdloss}
\mathcal{L}_\text{thr} = 
\underbrace{\mathcal{L}_\text{rec}}_{\text{clean}} + \lambda
\underbrace{\mathbb{E}_{\mathbf{x,n}}\!\left[\|D_{\boldsymbol{\phi}}(E_{\boldsymbol{\theta}}(\mathbf{x}_\text{trg})+\mathbf{n}) - \mathbf{x}_\text{tgt}\|^2\right]}_{\text{backdoor}},
\end{equation}
where $\lambda$ balances the two terms. 
Such a vulnerability has recently been demonstrated for deep JSCC
autoencoders, where a small trigger patch is sufficient to override
the deep JSCC model~\cite{sagduyu_backdoor,bass}.

At inference, the defender, located at the transmitter, has white-box
access to the backdoored pair
$(E_{\boldsymbol{\theta}},D_{\boldsymbol{\phi}})$ but no knowledge of
the trigger $\mathbf{p}$, trigger mask $\mathbf{m}$, or the target
image $\mathbf{x}_\text{tgt}$.
The defender aims to purify the input at inference, without
retraining $\boldsymbol{\theta}$ or $\boldsymbol{\phi}$.

\begin{figure*}
    \centering
    \includegraphics[width=1.0\linewidth]{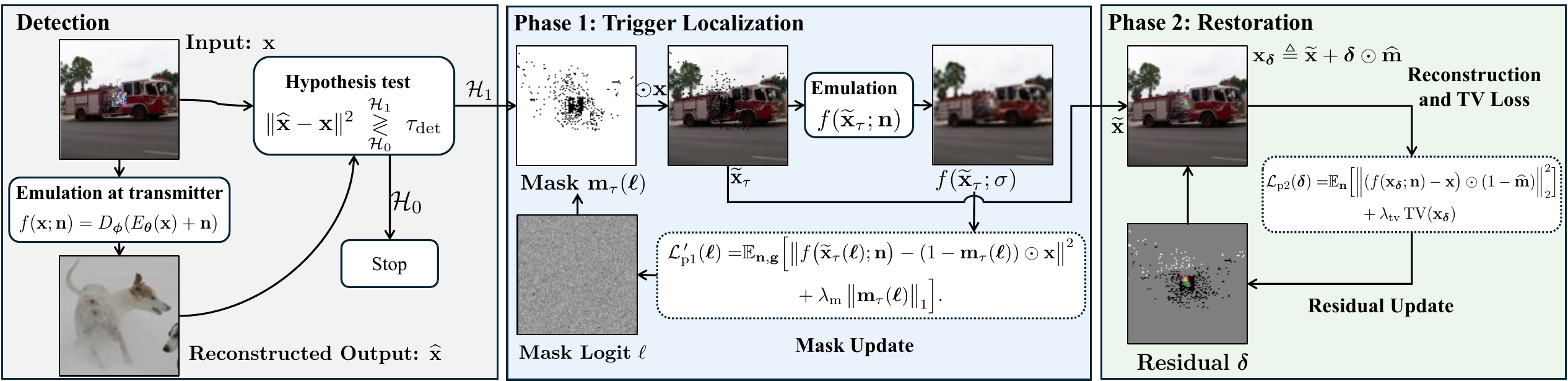}
    \caption{Overview of the proposed \MROP defense at the \textbf{transmitter}. A suspicious input flagged by the detector~\eqref{eq:detect} enters Phase 1, which localizes the trigger support $\widehat{\mathbf{m}}$ by optimizing a Gumbel-sigmoid mask through $\mathcal{P}_1'$. With $\widehat{\mathbf{m}}$ frozen, Phase 2 restores the trigger region via the residual $\boldsymbol{\delta}$ optimized through $\mathcal{P}_2$, after which the purified image is encoded, transmitted over the AWGN channel, and decoded.}
    \label{fig:MROP_concept}
\end{figure*}

\section{Proposed Defense Method: \MROP}

In this section, we introduce \MROP, a mask-region optimization method for backdoor trigger purification. 
The backdoor purification is previously addressed in~\cite{bass}; however, the defense scheme in~\cite{bass} requires a pair of trigger-injected images, which has a limitation in a practical situation. 
On the other hand, our method utilizes the property of the reconstruction task to detect and purify the triggers injected into input images.

\paragraph*{Detection}
Before introducing \MROP scheme, we introduce the detection policy for the adversarial patch, as depicted in Fig.~\ref{fig:MROP_concept}.
For notation brevity, we denote the end-to-end encoding and decoding process as a function $f:\mathbb{R}^n\rightarrow\mathbb{R}^n$, where $f(\mathbf{x};\mathbf{n}) = D_{\boldsymbol{\phi}}(E_{\boldsymbol{\theta}} (\mathbf{x})+\mathbf{n})$, which can be locally emulated prior to transmitting $\mathbf{z}$. 
This is a benign assumption: in a bidirectional system, where each terminal operates over both uplink and downlink, the encoder and decoder are co-located at every node by design.
Hence, the transmitter can decide whether the input image is trigger-injected or not by comparing the self-reconstruction error against a detection threshold $\tau_\text{det}$, i.e.,
\begin{equation}
\label{eq:detect}
\Vert f(\mathbf{x};\mathbf{n}) - \mathbf{x} \Vert^2
\;\underset{\mathcal{H}_0}{\overset{\mathcal{H}_1}{\gtrless}}\;
\tau_\text{det},
\end{equation}
where $\mathcal{H}_1$ and $\mathcal{H}_0$ denote the hypotheses that the input is trigger-injected and clean, respectively.

When the input is declared trigger-injected ($\mathcal{H}_1$), the transmitter invokes the proposed \MROP method; otherwise, the encoded message $\mathbf{z}=E_{\boldsymbol{\theta}}(\mathbf{x})$.


\paragraph*{Phase 1: Trigger localization}

In the first phase, the \MROP estimates the binary trigger mask by $\widehat{\mathbf{m}}\in\{0,1\}^n$.
With the estimated mask, we represent the masked composite of $\mathbf{x}$ by 
\begin{equation}
    \label{eq:mask_composite} 
    \tilde{\mathbf{x}}=(1-\widehat{\mathbf{m}})\odot \mathbf{x}.
\end{equation}
For the objective function of the binary mask optimization, we formulate a loss function as a weighted sum of reconstruction consistency and sparsity:
\begin{equation}
    \mathcal{L}_{\text{p1}}= \mathbb{E}_\mathbf{n}\left[\Vert f(\tilde{\mathbf{x}};\mathbf{n})-(1-\widehat{\mathbf{m}})\odot \mathbf{x}\Vert^2\right] + \lambda_\text{m}\,\Vert\widehat{\mathbf{m}}\Vert_0,
\end{equation}
where $\lambda_\text{m}$ denotes a weight factor, and the second term measures mask sparsity. 
Then the optimization is formulated by 
\begin{equation}
\label{eq:p1}
\tag{P1}
\mathcal{P}_1:\min_{\widehat{\mathbf{m}}}\mathcal{L}_{\text{p1}} \;\;\; \text{s.t. } \widehat{\mathbf{m}}\in\{0,1\}^n.
\end{equation}

Problem~\ref{eq:p1} is a binary optimization problem over $2^n$ candidate masks and is not amenable to gradient-based optimization owing to the binary constraint. 
To circumvent this combinatorial intractability, we relax the problem via the Gumbel-sigmoid trick~\cite{gumbel,concrete}: we attach a learnable logit vector
$\boldsymbol{\ell}\in\mathbb{R}^n$, where each entry $\ell_i$ governs the inclusion probability of pixel $i$ in the trigger mask. 
A differentiable surrogate of a binary mask sample is then drawn by
\begin{equation}
    \label{eq:bcm}
    \mathbf{m}_\tau(\boldsymbol{\ell})
    = \mathtt{sigmoid}\!\left(
        \frac{\boldsymbol{\ell} + \mathbf{g}}{\tau}
      \right),
\end{equation}
where $\mathtt{sigmoid}(\cdot)$ denotes the element-wise sigmoid function and $\mathbf{g}\in\mathbb{R}^n$ is the logistic noise vector whose entries are given by $g_i = \log u_i - \log(1-u_i)$ with $u_i\sim\mathrm{Uniform}(0,1)$.
The temperature $\tau>0$ controls the sharpness of the relaxation: as $\tau\to 0^{+}$, $\mathbf{m}_\tau(\boldsymbol{\ell})$ converges in distribution to a hard Bernoulli sample with parameter $\mathtt{sigmoid}(\boldsymbol{\ell})$, whereas a larger $\tau$ yields a smoother and more exploratory mask.
By construction, $\mathbf{m}_\tau(\boldsymbol{\ell})\in(0,1)^n$ is differentiable in $\boldsymbol{\ell}$, so the gradients flow back to the logit vector while the sample itself remains a faithful surrogate of the binary mask.

Substituting $\mathbf{m}_\tau(\boldsymbol{\ell})$ for $\widehat{\mathbf{m}}$ in $\mathcal{P}_1$, we define the relaxed masked composite by $\widetilde{\mathbf{x}}_\tau(\boldsymbol{\ell}) = (1-\mathbf{m}_\tau(\boldsymbol{\ell}))\odot \mathbf{x}$,
and, approximating the $\ell_0$ penalty by its $\ell_1$ counterpart, the relaxed loss function is given by
\begin{align} \label{eq:l_p1_prime}
\mathcal{L}_{\text{p1}}^{\,\prime}(\boldsymbol{\ell}) = & \mathbb{E}_\mathbf{n,g}\Big[\bigl\Vert f\bigl(\widetilde{\mathbf{x}}_\tau(\boldsymbol{\ell}); \mathbf{n}\bigr) 
- (1-\mathbf{m}_\tau(\boldsymbol{\ell}))\odot \mathbf{x}\bigr\Vert^2 \\
& ~~~~~~~~ + \lambda_\text{m}\,\bigl\Vert \mathbf{m}_\tau
(\boldsymbol{\ell})\bigr\Vert_1\Big]. \notag
\end{align}
The relaxed optimization is then formulated as 
\begin{equation}
\label{eq:p1r}
\tag{P1$'$}
\mathcal{P}_1^{\,\prime}:\min_{\boldsymbol{\ell}\in\mathbb{R}^n}\;
\!
\mathcal{L}_{\text{p1}}^{\,\prime}(\boldsymbol{\ell}),
\end{equation}
which is solved by $T_1$ iterations of stochastic gradient descent on $\boldsymbol{\ell}$. 
At step $t$, fresh logistic noise $\mathbf{g}^{(t)}$ is drawn and the logit vector is updated by
\begin{equation}
\label{eq:p1-update}
\boldsymbol{\ell}^{(t+1)} = \boldsymbol{\ell}^{(t)}
- \eta_1\,\nabla_{\boldsymbol{\ell}}\,
\mathcal{L}_{\text{p1}}^{\,\prime}
\bigl(\boldsymbol{\ell}^{(t)}\bigr),
\;\;\;
\tau_t = \tau_0
\!\left(\tfrac{\tau_{T_1}}{\tau_0}\right)^{\!t/T_1},
\end{equation}
where the temperature is exponentially annealed from $\tau_0$ to $\tau_{T_1}$ over the iterations. The annealing schedule lets the relaxation explore softly in early iterations and harden into a near-binary decision toward convergence. 


\paragraph*{Phase 2: Trigger-region restoration}

With the trigger $\widehat{\mathbf{m}}$ from Phase 1, the trigger region is localized but zeroed out in $\widetilde{\mathbf{x}}$, which is inconsistent with the surrounding pixels.
In Phase 2, we restore the content of the trigger region so that the decoded image becomes consistent with the non-trigger pixels.
To this end, we introduce a residual $\boldsymbol{\delta}\in\mathbb{R}^n$ confined to the trigger region, and optimize it so that the decoded output matches $\mathbf{x}$ in the non-trigger region.
To optimize the residual vector $\boldsymbol{\delta}$, the Phase-2 loss is defined as 
\begin{equation}
\label{eq:l_p2}
\mathcal{L}_{\text{p2}}(\boldsymbol{\delta})
=\mathbb{E}_{\mathbf{n}}\Big[\Big\Vert(f(\mathbf{x}_{\boldsymbol{\delta}};\mathbf{n})
-\mathbf{x}\bigr)\odot(1-\widehat{\mathbf{m}})\Big\Vert_2^2\Big]
+\lambda_\text{tv}\,\mathrm{TV}(\mathbf{x}_{\boldsymbol{\delta}}),
\end{equation}
where $\mathbf{x}_{\boldsymbol{\delta}}\triangleq\widetilde{\mathbf{x}}
+\boldsymbol{\delta}\odot\widehat{\mathbf{m}}$ is the Phase-2 input, $\lambda_\text{tv}$ is a weight, and $\mathrm{TV}(\cdot)$ is the isotropic total
variation~\cite{tv}.
The optimization is then formulated as
\begin{equation}
\label{eq:p2}
\tag{P2}
\mathcal{P}_2:\min_{\boldsymbol{\delta}}\;
\mathcal{L}_{\text{p2}}(\boldsymbol{\delta}).
\end{equation}
The mask $\widehat{\mathbf{m}}$ and the model $f$ are kept frozen, and only $\boldsymbol{\delta}$ is updated. 
The problem $\mathcal{P}_2$ is solved by $T_2$ iterations of gradient descent on $\boldsymbol{\delta}$,
\begin{equation}
\label{eq:p2-update}
\boldsymbol{\delta}^{(t+1)} = \boldsymbol{\delta}^{(t)}
- \eta_2\,\nabla_{\boldsymbol{\delta}}\,
\mathcal{L}_{\text{p2}}\bigl(\boldsymbol{\delta}^{(t)}\bigr),
\end{equation}
with step size $\eta_2$.
After convergence, the purified image $\widetilde{\mathbf{x}} +\boldsymbol{\delta}^\star\odot\widehat{\mathbf{m}}$ is encoded, transmitted through the channel, and decoded to obtain the final reconstruction.

The overall \MROP procedure---detection by~\eqref{eq:detect}, trigger localization by $\mathcal{P}_1'$, and trigger-region restoration by $\mathcal{P}_2$---requires no retraining of the JSCC model and operates entirely at inference.

\section{Numerical Results}

In this section, we evaluate \MROP scheme on two JSCC models: \textit{DeepJSCC}~\cite{deepjscc} and \textit{SwinJSCC}~\cite{yang2024swinjscc}, and two trigger patterns, and compare against an input-purification baseline adapted from~\cite{februus,sentinet}. 

\subsection{Simulation Setup}
\paragraph*{Dataset, model, and baseline} 
We evaluate \MROP on CIFAR-10 ($32\!\times\!32$)~\cite{cifar10} and the higher-resolution STL-10 ($96\!\times\!96$)~\cite{stl10}, over an AWGN channel at a fixed SNR of $19$ and $10$\,dB, respectively. 
Two JSCC models (DeepJSCC~\cite{deepjscc} and Transformer-based SwinJSCC~\cite{yang2024swinjscc}) are trained from scratch with the code rate of ($k/n=1/6$). 
The attacker poisons a fraction $0.2$ of the training data with a trigger patch and a fixed target image $\mathbf{x}_\text{tgt}$, minimizing~\eqref{eq:bdloss} with
$\lambda=1$.
We consider two trigger patterns: i)a center patch of size $\{4,6,8,10,12\}$\, pixels and ii) a universal patch at a random location.
For \MROP, Phase 1 anneals the temperature from $\tau_0=8$ to $\tau_{T_1}=0.1$ over $T_1=100$ iterations with $\lambda_\text{m}=0.02$, and Phase 2 refines $\boldsymbol{\delta}$ (initialized at zero) over $T_2=150$ Adam iterations; the baseline gradient-mask defense is swept over top-$\{5,15,25\}\%$ pixels.
Reported numbers are evaluated using the test sets of the datasets, averaged over up to five target images per
setting.

\paragraph*{Metrics} 
We report four metrics: i) the \emph{clean PSNR} (PSNR-C), measured between the reconstruction and the original image \emph{over the non-trigger region only}; ii) the \emph{clean SSIM} (SSIM-C); iii) the \emph{target PSNR} (PSNR-T), measured against $\mathbf{x}_\text{tgt}$ over the whole image, for which a large value indicates a successful attack; and iv) the \emph{attack success rate} (ASR), defined as the fraction of images whose reconstruction is closer to $\mathbf{x}_\text{tgt}$ than to $\mathbf{x}$ in the $\ell_2$ sense.

\begin{table}[t]\centering
\caption{Backdoor defense on CIFAR-10 (DeepJSCC) with a $4\!\times\!4$ patch, for the center and universal triggers. Best results in bold.}
\label{tab:main}
\adjustbox{width=1\linewidth}{
\begin{tabular}{llcccc}
\toprule
Trigger & Defense & PSNR-C & PSNR-T & SSIM-C & ASR (\%) \\
\midrule
\multirow{5}{*}{Center} & No defense & 9.66 & 47.72 & 0.059 & 99.99 \\
 & Gradient mask ($5\%$) & 10.80 & 42.08 & 0.119 & 93.24 \\
 & Gradient mask ($15\%$) & 13.44 & 31.26 & 0.278 & 70.66 \\
 & Gradient mask ($25\%$) & 14.70 & 24.54 & 0.362 & 52.76 \\
 & \textbf{MROP (ours)} & \textbf{28.43} & \textbf{9.78} & \textbf{0.889} & \textbf{0.00} \\
\midrule
\multirow{5}{*}{Universal} & No defense & 9.68 & 46.27 & 0.056 & 99.99 \\
 & Gradient mask ($5\%$) & 10.19 & 42.20 & 0.088 & 96.53 \\
 & Gradient mask ($15\%$) & 9.73 & 44.93 & 0.058 & 99.86 \\
 & Gradient mask ($25\%$) & 9.70 & 45.40 & 0.056 & 99.98 \\
 & \textbf{MROP (ours)} & \textbf{24.74} & \textbf{9.86} & \textbf{0.788} & \textbf{0.00} \\
\bottomrule
\end{tabular}
}
\end{table}

\begin{figure}
    \centering
    \includegraphics[width=0.85\linewidth]{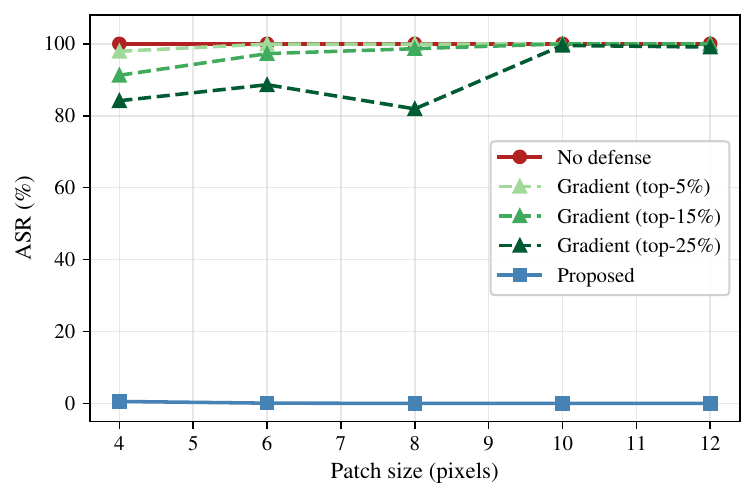}
    \vspace{-8pt}
    \caption{Attack success rate (ASR) versus trigger patch size for the center-patch attack on CIFAR-10 (DeepJSCC). }
    \label{fig:asr_vs_ps}
\end{figure}

\paragraph*{Comparison}
Table~\ref{tab:main} reports the defense performance on CIFAR-10 with a $4\!\times\!4$ patch. 
Without defense, both triggers attain an ASR of $99.99\%$ while the PSNR-C collapses to about $9.66$\,dB, confirming that the backdoor overrides the reconstruction almost everywhere. 
The gradient-mask baseline behaves very differently for the two triggers: for the center patch, its largest budget ($25\%$) lowers the ASR to $52.76\%$, but for the universal patch---whose location varies across images---it is essentially useless, leaving the ASR
above $99\%$ for every baseline, since a fixed saliency budget cannot track a moving trigger. 
In contrast, \MROP drives the ASR to $0.00\%$ for both triggers and raises the PSNR-C to $28.43$ and $24.74$\,dB, an improvement of more than $14$\,dB over the strongest baseline; the SSIM-C of $0.889$ and $0.788$ further confirms that the clean content is faithfully recovered. 
The low PSNR-T of \MROP (below $10$\,dB) indicates that the purified reconstruction no longer resembles the attacker's target.

\paragraph*{Robustness to patch size}
Figure~\ref{fig:asr_vs_ps} shows the ASR as the center patch grows from $4\times 4$ to $12\times 12$\,pixels on CIFAR-10. 
The gradient-mask baseline not only fails but also degrades as the patch enlarges, because a larger trigger forces a wider mask that erases too many clean pixels. 
\MROP, on the other hand, keeps the ASR below $2\%$ across the entire range, as its mask optimization is robust to changes in mask size.

\begin{table}[t]\centering
\caption{Experiment on SwinJSCC against center patch attack with CIFAR-10 ($32\!\times\!32$, $8\times8$\,pixels trigger) and STL-10 ($96\!\times\!96$, $12\times12$\,pixels trigger) datasets.}
\label{tab:swin}
\adjustbox{width=1\linewidth}{
\begin{tabular}{llcccc}
\toprule
Dataset & Defense & PSNR-C & PSNR-T & SSIM-C & ASR (\%) \\
\midrule
\multirow{5}{*}{\shortstack[l]{CIFAR-10}} & No defense & 9.18 & 47.53 & 0.067 & 99.99 \\
 & Gradient mask ($5\%$) & 9.21 & 44.37 & 0.070 & 99.89 \\
 & Gradient mask ($15\%$) & 9.50 & 38.75 & 0.088 & 98.63 \\
 & Gradient mask ($25\%$) & 11.38 & 31.82 & 0.181 & 81.90 \\
 & \textbf{MROP (ours)} & \textbf{28.96} & \textbf{9.34 }& \textbf{0.849} & \textbf{0.00} \\
\midrule
\multirow{5}{*}{\shortstack[l]{STL-10}} & No defense & 9.57 & 41.66 & 0.137 & 99.99 \\
 & Gradient mask ($5\%$) & 9.82 & 40.87 & 0.139 & 99.96 \\
 & Gradient mask ($15\%$) & 20.58 & 10.92 & 0.706 & 4.87 \\
 & Gradient mask ($25\%$) & 19.45 & 10.49 & 0.662 & \textbf{0.40} \\
 & \textbf{MROP (ours)} & \textbf{24.28} & \textbf{10.27} & \textbf{0.799} & 0.77 \\
\bottomrule
\end{tabular}
}
\end{table}

\paragraph*{Generalization across JSCC model and resolution}
Table~\ref{tab:swin} verifies that \MROP transfers directly to the Transformer-based SwinJSCC and to a higher resolution. 
On CIFAR-10 \MROP outperforms baselines with the ASR of $0.00\%$ at $28.96$\,dB PSNR-C, whereas the best baseline remains at $81.90\%$. 
On STL-10, the larger localizable trigger lets the gradient mask succeed ($0.4\%$ ASR at $25\%$), yet at a PSNR-C of only $19.45$\,dB; \MROP attains a comparable ASR of $0.77\%$ while delivering a $4.83$\,dB higher PSNR-C, i.e., a cleaner reconstruction at the same level of protection. 
These results show that \MROP generalizes across trigger type, patch size, network architecture, and image resolution.

\paragraph*{Qualitative results}

\begin{figure}
    \centering
    \includegraphics[width=0.95\linewidth]{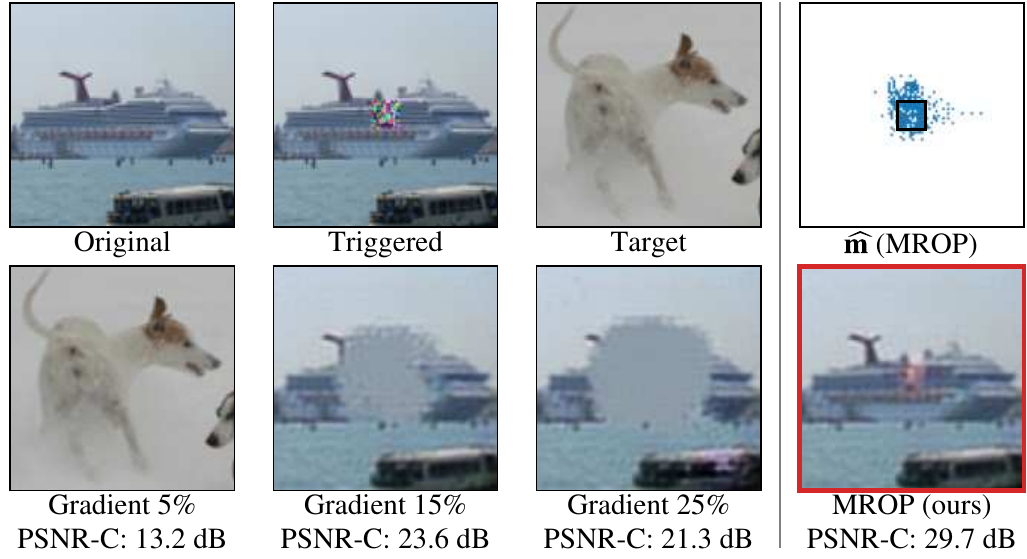}
    \caption{Qualitative comparison on a triggered STL-10 image (center patch). The fourth figure in the first row depicts the estimated and original trigger masks in \textit{blue} and \textit{black}, respectively.}
    \label{fig:qualitative}
\end{figure}

Figure~\ref{fig:qualitative} illustrates an STL-10 example: the recovered mask (blue) closely matches the ground-truth mask, and the resulting reconstruction is visually clean, whereas every baseline still fails to recover the shape of the original image.

\section{Conclusion}
We studied input-patch backdoor attacks on deep JSCC, where a small trigger forces the decoder to emit an attacker-chosen target image, and proposed \MROP, an inference-time purification method that requires no retraining of the JSCC model. 
Exploiting the reconstruction nature of the task, \MROP localizes the trigger by optimizing a per-pixel Gumbel-sigmoid mask and then restores the masked region under a total-variation prior so that the decoded image is consistent with the surrounding scene.  
Across two models, two trigger types, and a range of patch sizes and resolutions, \MROP suppressed the ASR to below $2\%$ while improving the clean-region PSNR by more than $10$\,dB over
a gradient-mask baseline. 


\bibliographystyle{IEEEtran}
\bibliography{main}

\end{document}